\documentclass[sigconf,natbib=false,nonacm=true]{acmart}

\AtBeginDocument{%
  }

\RequirePackage[
  datamodel=acmdatamodel,
  style=acmnumeric,
  ]{biblatex}

\begin{document}

\title{3DLNews: A Three-decade Dataset of US Local News Articles}

\author{Gangani Ariyarathne}
\orcid{0000-0003-4205-6574}
\affiliation{
  \institution{William \& Mary}
  \city{Williamsburg}
  \state{Virginia}
  \country{USA}
}
\email{gchewababarand@wm.edu}

\author{Alexander C. Nwala}
\orcid{0000-0003-3408-791X}
\affiliation{
  \institution{William \& Mary}
  \city{Williamsburg}
  \state{Virginia}
  \country{USA}
}
\email{acnwala@wm.edu}

\renewcommand{\shortauthors}{Gangani Ariyarathne \& Alexander C. Nwala}

\begin{abstract}
  We present \textit{3DLNews}, a novel dataset with local news articles from the United States spanning the period from 1996 to 2024. It contains almost 1 million URLs (with HTML text) from over 14,000 local newspapers, TV, and radio stations across all 50 states, and provides a broad snapshot of the US local news landscape. The dataset was collected by scraping Google and Twitter search results. We employed a multi-step filtering process to remove non-news article links and enriched the dataset with metadata such as the names and geo-coordinates of the source news media organizations, article publication dates, etc. Furthermore, we demonstrated the utility of 3DLNews by outlining four applications.
\end{abstract}


\maketitle

\section{Introduction}

With over 329 million Americans across 3,143 counties, the national media alone cannot provide news coverage for every community. Thus, local media plays a critical role by focusing on local issues, including government waste, corruption, the effectiveness of public schools, etc. In fact, more than half of original news content is produced by local media~\cite{Mary_Ellen_Klas_local_news_democracy}. Local media was responsible for leading reports for many important stories, including exploring how the \textit{Opioid epidemic} destroyed many lives in McDowell, West Virginia \cite{SPerry_opioid}, chronicling the \textit{Flint water crisis} before it received the national spotlight \cite{DRobbins_flint}, and more recently reporting on the multi-faceted experiences of various communities experiencing the COVID-19 pandemic~\cite{Elisa_Shearer_covid_local_news}. Given the significance of local media, local news datasets are crucial for studying the US and investigating various conditions experienced by residents of small towns and cities, surrounding issues of health, democracy, economy, etc.

Existing news article datasets~\cite{horne2022nela,roberts2021media,weaver2008finding,norregaard2019nela,leetaru2013gdelt} either focus on global or national news, are paywalled, or are limited in scope; covering only specific geographical areas, timeframes, or topics. In response to this gap, we present \textit{3DLNews}, the first-ever collection of US local news articles published from 1996 to 2024. 3DLNews was collected by scraping Google and Twitter (now X) search results. It contains about 1 million links of US local news articles from more than 14,000 websites of local newspapers, TV, and radio stations across all 50 states. The extracted URLs were carefully filtered to remove non-news article links. Furthermore, we enriched the dataset by including attributes such as, the names and geo-coordinates of the source news media, article publication dates, HTML text, etc. We published both the filtered and original dataset~\cite{gangani_nwala_3dlnews}. We demonstrated the usefulness of 3DLNews by outlining four use cases including, exploring the nationalization of local news, media bias and local news desert analyses, and community understanding.

\section{Related Datasets}
In contrast with existing news datasets, 3DLNews is focused on US local news, free, covers all 50 US states from 1996 to 2024, and features news articles across a wide variety of topics. Table~\ref{tab:comparison} summarizes the similarities and differences between 3DLNews and the existing news datasets discussed next.

\textbf{Media Cloud} \cite{roberts2021media} (\textcolor{blue}{\url{www.mediacloud.org}}), is an open-source web platform with news articles from thousands of global and national news outlets, blogs, etc. In addition to the news dataset, it offers web services such as, analytical tools for topic, media bias, network analysis, and trend tracking.

\textbf{LexisNexis} \cite{weaver2008finding} (\textcolor{blue}{\url{www.lexisnexis.com}}), is a commercial database with news articles, legal documents, business information, etc. Even though LexisNexis has a global focus, we do not know the quantity of US local news articles in the dataset since it is not free to access.

\textbf{Nela-GT}~\cite{norregaard2019nela, gruppi2020nelagt2019, gruppi2021nela, gruppi2023nelagt2022}, the News Ecosystem Learning Agent-Ground Truth (Nela-GT) dataset, is a labeled global news dataset for studying misinformation in news articles.

\textbf{GDELT} \cite{leetaru2013gdelt} (\textcolor{blue}{\url{www.gdeltproject.org}}), the Global Database of Events, Language, and Tone (GDELT) dataset, is as a database that tracks broadcast, print, and web news in over 100 languages. It includes semantic labels for entities such as names of people, organizations, locations, events, etc. In addition to data, GDELT offers an online service for analyzing global societal trends.

\textbf{NELA-Local} \cite{horne2022nela} is a collection of over 1.4 million US local news articles collected from 313 local news outlets within 20 months (April 4, 2020 -- December 31, 2021). Each article includes metadata associated with the community served such as, county demographics, politics, etc. Even though NELA-Local is most similar to 3DLNews since they both focus on US local news articles across multiple topics, there are significant differences. First, 3DLNews covers a 28-year period (vs. the 20 months). Second, 3DLNews includes news articles from 14,086 news outlets (vs. 313). Third, unlike 3DLNews, NELA-Local is a longitudinal dataset which was created by extracting news articles daily from RSS feeds.

\begin{table}[H]
\caption{Comparison of existing news datasets and 3DLNews.}
\label{tab:comparison}
\begin{tabular}{lccc} 
\toprule
Dataset & Time range & \# Articles & Free/Paid\\ 
\midrule
Media Cloud \cite{roberts2021media} & 2008 -- Present & $\sim$1.7 billion & Free\\
Lexis Nexis \cite{weaver2008finding} & 1980 -- Present & 83 billion & Paid \\
Nela-GT \cite{norregaard2019nela} & 2018 -- 2022 & $\sim$7 million & Free \\
GDELT  \cite{leetaru2013gdelt}  & 1979 -- Present &  $\sim$6 million & Free \\
NELA-Local \cite{horne2022nela} & 2020 -- 2021 & 1.4 million & Free \\
\textbf{3DLNews \cite{gangani_nwala_3dlnews}} & \textbf{1996 -- 2024} & \textbf{$\sim$1 million} & \textbf{Free} \\ 
\bottomrule
\end{tabular}
\end{table}

\section{Building 3DLNews}

Here we explain our steps for creating 3DLNews.

\subsection{Local news media dataset}

We used an extended version of the Local Memory Project's ~\cite{jcdl-2017:nwala:lmp} (LMP) US local news dataset as seeds for our data extraction. LMP's dataset consists of the websites of 5,993 local newspapers, 2,539 TV stations, and 1,061 radio stations, primarily extracted from \textcolor{blue}{~\url{thepaperboy.com}} in 2016. We extended it by crawling and scraping\textcolor{blue}{~\url{thepaperboy.com}} (again), \textcolor{blue}{~\url{web.archive.org/web/20221203031956/http://www.usnpl.com/}}, \textcolor{blue}{\url{50states.com}} (similar to Nela-Local~\cite{horne2022nela}), and \textcolor{blue}{\url{einpresswire.com/world-media-directory/3/united-states}}. This resulted in the publicly released local news dataset~\cite{gangani_nwala_3dlnews} outlined in Table~\ref{tab:news_websites}. The ``broadcast'' type refers to either TV or radio stations, because we could not accurately distinguish them during scraping.

\subsection{Data extraction and filtering}

\textbf{Step 1:} We created Google search queries for each website in the local news media dataset (Table ~\ref{tab:news_websites}). 
For a single media website, e.g., \textcolor{blue}{\url{timesstar.com}}, we constructed the following Google query: 
\begin{verbatim}
https://www.google.com/search?tbs=cdr:1,
cd_min:1/1/1996,cd_max:12/31/1996
&q=news%20site:http://www.timesstar.com/.
\end{verbatim}
This instructs Google to return ``news'' webpages exclusively from \textcolor{blue}{\url{timesstar.com}} and published in 1996. We changed the Google date directives (\texttt{cd\_min}/\texttt{cd\_min}) to retrieve webpages published in a different year. Next, similar to Google, for each website, e.g., \textcolor{blue}{\url{timesstar.com}} in the local news media dataset, we created Twitter search queries as follows:
\begin{verbatim}
'timesstar.com' until:2006-12-31 since:2006-01-01
\end{verbatim} 
This instructs Twitter to return tweets posted in 2006 and linked to \textcolor{blue}{\url{timesstar.com}}. We changed the Twitter date directives (\texttt{until}/\texttt{since}) to retrieve webpages published in a different year.

The goal of the 3DLNews dataset is to provide a representative sample of US local news stories from 1996 to 2024, rather than capturing every published article. Therefore, our Google scraping focuses solely on the first page of results, and our Twitter scraping includes only the top 20 tweets per query.

\textbf{Step 2:} We issued Google and Twitter search queries to their respective search engines and scraped their links. For Google, we created queries from 1996 -- 2024, for Twitter, 2006 -- 2024. Table~\ref{tab:3dlnews_combined} presents the number of links scraped from Google and Twitter for each media type. It includes both news and non-news article links (e.g., homepages, menu pages) and corresponding counts for successfully downloaded HTML files. We removed non-news links by applying a filtering process outlined in Step 3. Table~\ref{tab:3dlnews_combined} presents the number of news articles (and HTML files) after filtering. 

\begin{table}
\centering
\caption{US local news media dataset.}
\label{tab:news_websites}
\begin{tabular}{cc}
\hline
Media Type & Number of websites \\ \hline
Newspapers & 9,441\\
Radio      & 2,449 \\
Broadcast  & 1,310  \\
TV         &  886 \\ 
\textbf{Total} & \textbf{14,086}  \\ \hline
\end{tabular}
\end{table}

\begin{table}
\caption{3DLNews: Counts of URLs collected (All URLs) and News article URLs after filtering applied (News URLs), and the counts of HTML files downloaded for All URLs (All HTML) and News URLs (News HTML).}
\label{tab:3dlnews_combined}
\begingroup
\begin{tabular}{lccccc} 
\toprule
       & All  & All  & News  & News    \\ 
Media       &  URLs &  HTML &  URLs &  HTML   \\ 
\midrule
          &   & \textbf{Google} &  &    \\
        \midrule
        Newspaper  & 853,543  & 636,967 & 502,530 & 436,031    \\
               Radio       & 140,401  & 113,383 & 52,925 & 52,117     \\
               TV          & 99,001   & 88,620 & 62,727 & 59,609     \\
               Broadcast   & 164,028  & 155,445 & 110,494 & 107,439    \\
              \midrule
                        &   & \textbf{Twitter} &  &    \\
        \midrule
        Newspaper  & 199,996  & 155,083 & 54,295 & 30,981    \\
              Radio       & 102,494  & 41,917 & 3,794 & 2,637       \\
              TV          & 66,880   & 54,632 & 13,213 & 5,895     \\
              Broadcast   & 100,119  & 44,909 & 10,497 & 7,921      \\
             \midrule
\textbf{Total}  & \textbf{1,727,462} & \textbf{1,290,956} & \textbf{810,475} & \textbf{702,630} \\ \bottomrule
\end{tabular}
\endgroup
\end{table}

\begin{figure*}
  \centering
  \includegraphics[width=\linewidth]{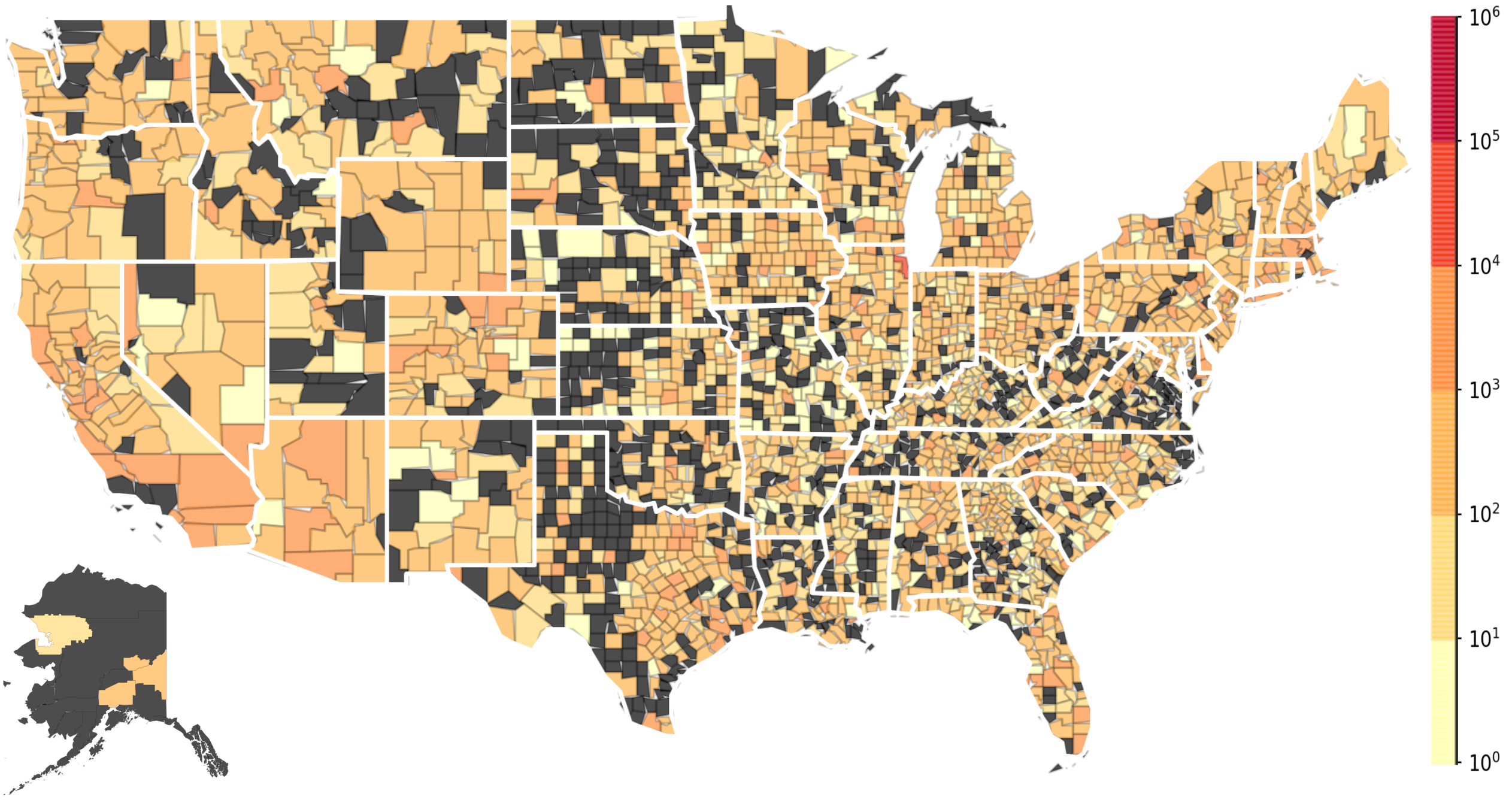}
  \caption{Local news articles in 3DLNews per US county. Black-colored counties indicate areas without news articles in 3DLNews.}
  \Description{Figure shows the distribution of local news articles in 3DLNews per US county. Black-colored counties indicate areas without news articles in 3DLNews. 3DLNews covers 100\% (50/50) of US states and about 68\% (2,146/3,143) of all US counties. Cook County, Illinois, the second-most-populous county in the US.}
  \label{fig:us-county-map}
\end{figure*}

Next, we outline our filtering process for removing non-news article URLs from 3DLNews (Table~\ref{tab:3dlnews_combined}), since there is no standard URL format for news articles.  We provided access to the raw data, enabling researchers to apply their own filtering. This process was informed by an experiment in which we created a gold-standard dataset of news article URLs to understand two properties: path depth and word-boundary. The path depth of a URL is the number of hierarchies in the URL path property. For example, \textcolor{blue}{\url{https://example.com/}} has a path depth of zero while \textcolor{blue}{\url{https://example.com/foo}} and \textcolor{blue}{\url{https://example.com/foo/bar}} have path depths of one and two, respectively. A word-boundary is simply a symbol that separates words in a URL. For example, the word-boundary for the URL, \textcolor{blue}{\url{https://example.com/this-is-a-page}} is `-'.

\textbf{Step 3:} First, we dereferenced all URLs to resolve redirects and retrieved final URLs that returned HTTP 200 codes. Second, we removed links with domains not present in our local news media dataset. Third, we lowercased all URLs, discarded trailing slashes, and removed duplicate URLs. Fourth, URLs with a path depth of zero, typically representing homepages, were removed since we only care about news article URLs which occur at a deeper path depth (e.g., $\ge$ 3). On some occasions however, we observed that news URLs occurred at lower path depths (e.g., $<3$). Therefore, we kept such news article URLs only if they included popular word-boundary separators such as `-', `\_', or `.' (e.g., \textcolor{blue}{\url{http://kwgs.org/post/funeral-set-ou-quarterback-killed-crash}}). We kept all URLs with path depth $\ge$ 3. Our 3DLNews dataset ~\cite{gangani_nwala_3dlnews} includes both raw and filtered versions with HTML text whenever available (Table~\ref{tab:3dlnews_combined}).

\subsection{Data Enrichment and format}
We enhanced the usefulness of the news article URLs in 3DLNews by adding attributes to each URL. See Table~\ref{tab:news_properties} for the complete list of attributes. Next, we highlight a few.

\textit{link} represents the news article URL. The \textit{html\_filename} attribute points to the file containing the HTML text of the news article, while the \textit{publication\_date} refers to the article publication date which was extracted using htmldate~\cite{barbaresi-2020-htmldate}. The \textit{location} property of each URL includes the US state, city, and latitude/longitude of the source news media organization. \textit{media\_metadata} contains information (e.g., newspaper or TV or radio name) about the news media website where the article was published. \textit{source\_metadata} includes information (e.g., search query link) about the source (Twitter or Google) from which the article was scraped.

Each news article URL, along with its attributes (Table ~\ref{tab:news_properties}) is encapsulated in a JSON object within a single line in a file in 3DLNews.

\section{Data Coverage and Descriptive Statistics}

Here we summarize the composition of 3DLNews along the geographic and temporal dimensions.

\subsection{Location Based Analysis}
Figure~\ref{fig:us-county-map} is a choropleth map illustrating the distribution of the counts of news articles per US county. Accordingly, 3DLNews covers 100\% (50/50) of US states and about 68\% (2,146/3,143) of all US counties. Cook county, Illinois, the second-most-populous county in the US, had the most news articles (13,006). In contrast, there were 59 counties with a single news article. The wide coverage of 3DLNews allows for robust analysis of regional news trends, community-specific issues, and the overall US local news ecosystem. Figure~\ref{fig:us-county-map} also reveals the presence of news deserts (black-colored counties); areas with no local news coverage. This could be attributed to the absence of local news media organizations in these counties or blind spots in 3DLNews. Therefore, further research is needed to determine the actual cause.

\begin{table}[H]
  \caption{Properties of news article URLs in 3DLNews.}
  \label{tab:news_properties}
  \begin{tabular}{p{2.2cm}p{5.6cm}} 
    \toprule
    Property & Description \\
    \midrule
    link                & The URL of the local news article.         \\
    html\_filename      & Filename with HTML content of the article. \\
    publication\_date   & Article publication date.                  \\
    title               & Title of the article.                      \\
    media\_name         & Name of local media organization.          \\
    media\_type         & Type of media source (\textit{Newspaper} or \textit{TV} or \textit{Radio station} or \textit{Broadcast}). ``Broadcast'' refers to either TV or radio stations. \\
    location            & Location of the media organization. This includes: US state, city, \& latitude/longitude.  \\
    media\_metadata     & More information about the news media.    \\
    source              & Platform (Google or Twitter) where the news article was extracted from. \\
    source\_metadata    & More information about the platform. \\
    response\_code      & Response code from issuing GET on link. \\
    expanded\_url       & Final target URL for links that redirect. \\
    \bottomrule
  \end{tabular}
\end{table}

\subsection{Time Based Analysis}

\begin{figure}
  \centering
  \includegraphics[width=\linewidth]{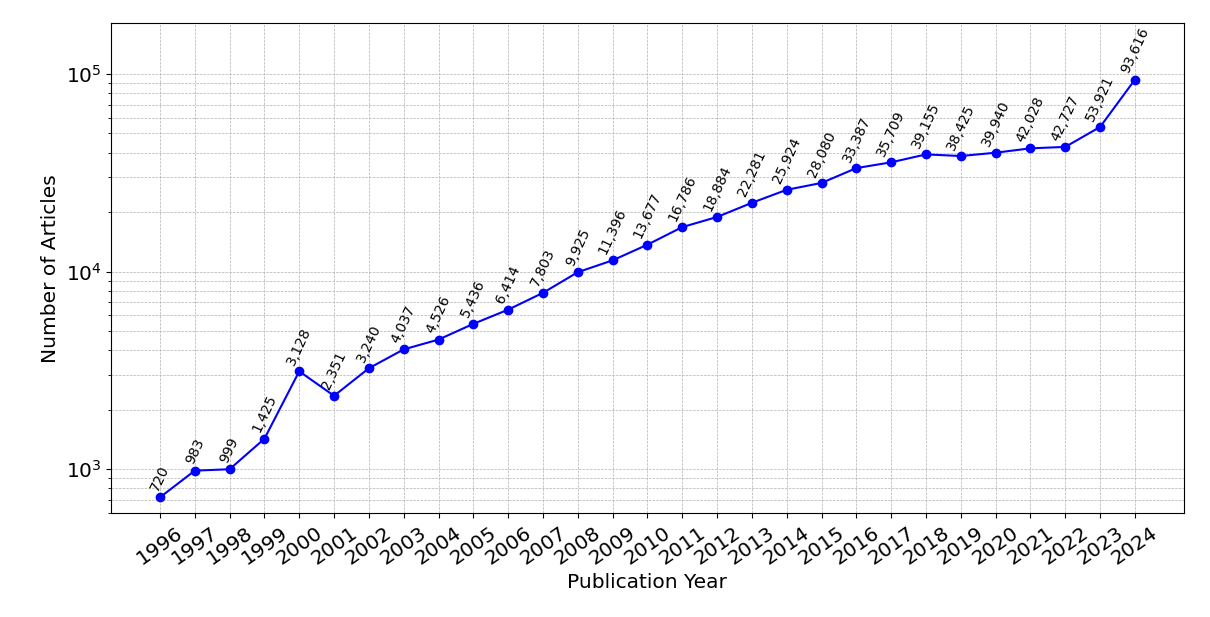} 
  \caption{Counts of news articles in 3DLNews per year.}
  \label{fig:time-based-analysis}
  \Description{Figure shows the distribution of the counts of local news articles from 1996 to 2024. The number of news articles gradually increases over time}
\end{figure}

Figure~\ref{fig:time-based-analysis} depicts the distribution of the counts of local news articles from 1996 to 2024. Overall, the number of news articles gradually increases over time. This upward trend highlights the growing volume of online local news content over the years. In the earlier years of our dataset, fewer articles were available, which is indicative of the developing stage of digital news and the limited web presence of local news outlets during that period. 

\section{Use Cases}

Here we demonstrate the utility of 3DLNews by outlining four possible applications that we intend to implement in future research.

\subsection{Exploring the Nationalization of Local News}
Local news outlets are meant to prioritize local stories that are relevant to their communities. This includes coverage of local politics, economy, culture, etc. A concerning contributor to the decline of local media in the US has been a steady increase in the nationalization of local news~\cite{bradshaw2019nationalization, jaidka2023news} --- the prioritization of national news (especially politics) over local news. 3DLNews presents a unique opportunity to quantify the degree of nationalization of local news across the US. This would involve using various natural language processing techniques to compare the topics of the news articles in 3DLNews with national news articles published in the same period.

\subsection{Media Bias Analysis}
Similar to the problem of the nationalization of local news, media bias in local news is another concerning issue that warrants further investigation, especially since bias could undermine the trust that local media enjoys over national media~\cite{knight_media_trust}. 3DLNews provides a comprehensive snapshot of the US local media landscape for studying bias in news coverage.

\subsection{Studying US Local News Deserts}
For the last two decades, thousands of local news stations have closed resulting in "news deserts" --- communities without a local news outlet --- spread across the US. In fact, since 2005, the US has lost almost 2,900 local newspapers and they continue to vanish at an average rate of more than two a week~\cite{Penelope_Muse_Abernathy_local_news_state}. 3DLNews may be used to study this phenomenon through the analyses of the density of local news articles relative to various geographic regions.

\subsection{Community Understanding} 
Given the broad geographical scope of 3DLNews, through content analysis, researchers can study the US through the lens of local media to gain deeper insights into living conditions in various communities or community attitudes surrounding various political, health, or economic issues.

The three-decade span of 3DLNews also provides a broad temporal scope for researchers to examine trends in local news coverage, to identify emerging topics, monitor changing public concerns, and to anticipate where the news agendas may be heading in the future. The insights gained from trend analysis and prediction have practical applications for journalists and possibly policymakers.

\section{Discussion}

Despite the broad geographic and temporal scope, and potential applications of 3DLNews, it has some limitations. 

First, while our filtering attempts to minimize the likelihood of including non-news article URLs, since there is no standard URL format of news articles, we expect to have included some small proportion of non-news article URLs in 3DLNews. To address this and other limitations of 3DLNews, we provided access to the raw data (URLs, HTML text, etc), enabling researchers to apply their own filtering and/or research-specific analyses. 

Second, 3DLNews excludes archived URLs (if they exist) for unavailable news articles. We plan to address this issue in future updates. Third, we were constrained by web scraping which limited the number of URLs we could collect.

Fourth, it is likely that 3DLNews includes articles from closed news organizations, since many news organizations have shut down between 1996 and 2024. Also, it is possible that 3DLNews does not include articles from closed news organizations whose information have been purged from search engine indexes. In a future research effort, we will quantify the proportion of articles for either cases and assess their impact. Furthermore, we intend to utilize the Internet Archive to understand how well local news articles are preserved.

Fifth, in our location-based analysis of news articles, we relied on the locations of the news media organizations, which might not always reflect the actual geographic areas covered by news stories. To remedy this, for future work, we will identify the specific geographic regions referenced in news stories.

\section{Conclusion}

We presented 3DLNews, a novel dataset with local news articles from the United States spanning the period from 1996 to 2024. 3DLNews is a significant contribution in the study of US local news coverage due to its broad geographical (covering 50 US states) and temporal scope (three-decade span). The dataset contains nearly 1 million URLs, enriched HTML text, and additional metadata for over 14,000 local news outlets across all 50 states and 68\% of US counties. We believe that 3DLNews provides a valuable opportunity for researchers to study the US and/or US local media ecosystem.

\begin{acks}
We thank the NSF for funding this work (award no. 2245508). The NSF had no role in designing our study, data collection and analysis, decision to publish, or preparation of the manuscript.
\end{acks}

\newpage
\printbibliography
\end{document}